\documentstyle[12pt]{article}
\begin{document}

\author{R. Vilela Mendes \\
Grupo de F\'\i sica-Matem\'atica\\
Complexo II, Universidade de Lisboa \\
Av. Gama Pinto, 2, 1699 Lisboa Codex Portugal\\
e-mail: vilela@alf4.cii.fc.ul.pt}
\title{Collision states and scar effects in charged three-body problems}
\date{}
\maketitle

\begin{abstract}
Semiclassical methods form a bridge between classical systems and their
quantum counterparts. An interesting phenomenon discovered in this
connection is the scar effect, whereby energy eigenstates display
enhancement structures resembling the path of unstable periodic orbits.

This paper deals with collision states in charged three-body problems, in
periodic media, which are scarred by unstable classical orbits. The scar
effect has a potential for practical applications because orbits
corresponding to zero measure classical configurations may be reached and
stabilized by resonant excitation. It may be used, for example, to induce
reactions that are favoured by unstable configurations. 
\end{abstract}

\section{Introduction}

Classical and quantum mechanics have qualitatively different features.
Nevertheless, semiclassical methods\cite{Maslov} provide, in some restricted
domains, a bridge between the quantum and classical realms. The Van
Vleck-Gutzwiller\cite{Van Vleck} \cite{Gutzwiller1} propagator and the trace
formula\cite{Gutzwiller2} are landmarks in this connection.

The trace formula relates the fluctuating part of the quantum density of
states to an oscillatory sum of exponentials, each term corresponding to a
classical periodic orbit or to one of its multiple retracings. The actual
convergence of the periodic orbit sum to the quantum spectrum is still an
open question\cite{Voros}. Nevertheless many interesting results have been
obtained by extracting quantum eigenvalue information from the classical
periodic orbits\cite{Periorbit}. 

In the trace formula one must sum over all periodic orbits to find a single
eigenvalue and in Van Vleck-Gutzwiller propagator $G(q,q^{^{\prime }},t)$,
the momentum uncertainty of the $q$ states implies that trajectories of all
energies must be taken into account. This led some authors\cite{Tomsovic} to
suggest that the reliability of the semiclassical results would be improved
if one propagates smooth square-integrable wave functions with finite-energy
uncertainty. The underlying idea is that the phase-space localized wave
functions would filter out the relevant information from the Green's
function.

The most interesting wave function information obtained from semiclassical
considerations is probably the {\it scar effect}\cite{Heller}. For complex
systems having phase-space domains with sensitive dependence to initial
conditions, periodic orbits are unstable and even when dense they are
nevertheless a zero measure set in the smooth ergodic measure. Naively, one
would then expect a typical wave function to receive contributions from many
orbits and its intensity to reflect the statistical average of them all. If
however the intensity of the wave function happens to be either concentrated
(or abnormally weak) near a classical periodic orbit one says that {\it the
quantum state is scarred by the periodic orbit}. The occurrence of this
behavior is easily understood from a wavepacket propagation argument\cite
{Heller} (see below). It has been observed in numerical calculations of
quantum eigenstates of several classically chaotic systems\cite{Heller} \cite
{McDonald} \cite{Taylor} \cite{Saraceno} and was related to unstable
periodic orbits through semiclassical path integrals\cite{Bogomolny} \cite
{Berry} \cite{Feingold}.

The scar effect may have far-reaching implications for the practical
applications of quantum systems. Namely, the unstable classical orbits in
chaotic systems , even if dense in phase space, are in practice never
observed because all typical motions are aperiodic and uniformly reproduce
the support of some diffuse invariant measure. By contrast, in quantum
mechanics, whenever an unstable periodic orbit scars a quantum energy
eigenstate, the system may easily be made to behave like the unstable orbit
by resonant excitation to the corresponding energy level. In this sense
scars are a gift of Nature, because they allow the exploration of dynamical
configurations that in classical mechanics are washed away by ergodicity. An
interesting scar effect has recently been observed\cite{Wilkinson} on a
semiconductor quantum-well tunneling experiment where, by localizing the
probability density, the scarring of the quantum well states increases the
overlap with the emitter and leads to enhanced tunnelling for some voltage
values.

In this paper I will be concerned with the relevance of the scar effect in
accessing collision or near-collision configurations of charged particles in
periodic media. Of particular interest are scars associated to separatrix
orbits of which a special type, here called {\it the saddle point scar}, is
an example. A simple wave-packet propagation argument\cite{Heller}
qualitatively explains the scar effect. It also allows the derivation of
simple conditions which will be important in the interpretation of the scar
effects treated in this paper. Consider the overlap integral 
\begin{equation}
\label{1.1}C(t)=\left\langle \Psi (t,x)\mid \Psi (0,x)\right\rangle 
\end{equation}
for a propagating wave packet which at time zero has a Gaussian shape and
initial conditions $(p_0,x_0)$ corresponding to an unstable periodic orbit.
Expanding $\Psi (0,x)$ in energy eigenstates 
\begin{equation}
\label{1.2}\Psi (0,x)=\sum_nc_n\Psi _n(x)
\end{equation}
one sees that the Fourier transform $S(E)$ of the overlap $C(t)$ is the
spectral density weighted by the probabilities $\left| c_n\right| ^2$. 
\begin{equation}
\label{1.3}S(E)=\sum_n\left| c_n\right| ^2\delta (E-E_n)
\end{equation}
On the other hand if the period $\tau $ of the classical periodic orbit and
the largest positive Lyapunov exponent $\lambda $ are such that $e^{-\tau
\lambda /2}$ is not too small, the overlap $C(t)$ will display peaks at
times $n\tau $. As the wave packet spreads, the amplitude of the peaks
decreases after each orbit traversal at the rate $e^{-\tau \lambda /2}$. The
Fourier transform of $C(t)$ will therefore have peaks of width $\lambda $
with spacing $\omega =\frac{2\pi }\tau $. From Eq.(\ref{1.3}) one concludes
that only the eigenstates that lie under the peaks contribute to the
expansion of the wave packet. Since the wave packet has an enhanced
intensity along the region of the period orbit, this is expected to carry
over to the contributing energy eigenstates. This is the {\it scar effect}.
The stronger the overlap resurgences are, the stronger the effect is
expected to be. Therefore the intensity of the effect varies like $1/\tau
\lambda $ .

This simple derivation\cite{Heller} of the scar effect is however flawed if
the product $\lambda d(E)$ (where $d(E)$ is the mean level density) is very
large. Then the number of contributing eigenstates is very large and no
individual eigenstate is required to show a significant intensity
enhancement near the periodic orbit. Also the argument assumes the low
period unstable orbits to be isolated. With many nearby orbits having
different periods the argument also breaks down. Conversely, if it happens
that in the same configuration space region many different periodic orbits
coexist, having the same period, the effect will be enhanced. This is the
situation for periodic motions in the neighborhood of an unstable critical
point (a saddle) of a potential function $V(x)\,$. Generically, in the
neighborhood of a critical point, there are coordinates where the function
may be written as 
\begin{equation}
\label{1.4}V(x)=\sum_i\frac 12\sigma _ix_i^2
\end{equation}
On the neighborhood of the point $x_i=0$ there are harmonic periodic motions
along the stable directions (positive $\sigma $'s) of the critical point. As
long as anharmonic corrections are unimportant all the orbits have the same
period independently of their amplitude. The instability parameter of these
orbits is the smallest negative $\sigma $ ($\lambda =-|\sigma _{\min }|$).
For each positive $\sigma _i$ the scar intensity factor will be $\frac
1{\tau \lambda }=\frac 1{2\pi }\sqrt{\frac{\sigma _i}\mu }\frac 1{|\sigma
_{\min }|}$ , where $\mu $ is the effective mass. The motion in the
neighborhood of these unstable periodic orbits being harmonic, the energy of
the strongest scar is estimated to be 
\begin{equation}
\label{1.5}V(0)+E_{loc}+\frac \hbar 2\sqrt{\frac{\sigma _{\max }}\mu }
\end{equation}
where the last term corresponds to the ground state energy of the harmonic
motion on the stable direction and $E_{loc}$ is the energy penalty
corresponding to the localization energy in the transversal directions. In
favorable conditions, that is, if the quadratic approximation for the
potential holds over a sufficiently large range, higher energy scars might
also be observed corresponding to the excited states of this oscillator.
They would have the approximate energy 
\begin{equation}
\label{1.5a}V(0)+E_{loc}+\left( n+\frac 12\right) \hbar \sqrt{\frac{\sigma
_{\max }}\mu }
\end{equation}

In the following two sections we study a one-dimensional and a
three-dimensional problem where unstable classical motions indeed leave
their trace in the some of the quantum states. Of particular interest here
are the unstable classical motions which correspond to collision or
almost-collision configurations. Some of the potential practical
applications of this effect are discussed in the conclusions.

\section{Scarred collision states in a one-dimensional periodic problem}

Let a quantum system be described by the Hamiltonian 
\begin{equation}
\label{2.1}H_1=-\frac{\partial ^2}{\partial x_1^2}-\frac{\partial ^2}{%
\partial x_2^2}-\frac 1\gamma \frac{\partial ^2}{\partial y^2}%
+V(|x_1-x_2|)-V(|x_1-y|)-V(|x_2-y|)
\end{equation}
with periodic boundary conditions on a lattice of lattice size $L$. The
Hamiltonian represents a many-body system with two heavy and one light
particle ($\gamma $ of order $10^{-4}$, for example) in each lattice cell
and periodic boundary conditions or, alternatively, a three-particle system
living on the circle. The heavy particles repel each other and attract the
light one. For definiteness I have considered the heavy particles to be
positively charged and the light particle to be the negative one. The
interaction potential is 
\begin{equation}
\label{2.2}V(x)=\frac g{2L}\left( 1+\cos (2\pi \frac xL)\right) 
\end{equation}
The factor $\frac 1L$ is included in the coupling constant $\frac gL$ as a
convenient factorization in case one wants to insure scaling properties
similar to the three-dimensional Coulomb problem. Periodic boundary
conditions are imposed on the quantum problem by choosing a basis of
box-normalized momentum eigenstates with periodic boundary conditions $\frac
1{\sqrt{L}}\exp (i2\pi kx)$ , $k=\frac nL$ , $n=0,\pm 1,\pm 2,\cdots $.
Denote by$\left| n_1n_2p\right\rangle $ a state with momenta $\frac{n_1}L$ , 
$\frac{n_2}L$ and $\frac pL$ respectively. The matrix elements of $H_1$ in
this basis are
\begin{equation}
\label{2.3}
\begin{array}{ccl}
\left\langle n_1^{^{\prime }}n_2^{^{\prime }}p^{^{\prime }}\left| H_1\right|
n_1n_2p\right\rangle  & = & \delta _{n_1^{^{\prime }}n_1}\delta
_{n_2^{^{\prime }}n_2}\delta _{p^{^{\prime }}p}(2\pi )^2\frac 1{L^2}\left(
n_1^2+n_2^2+
\frac{p^2}\gamma \right)  \\  &  & +\frac gL\delta _{pp}\delta
(n_1-n_1^{^{\prime }}+n_2-n_2^{^{\prime }})f_1(n_1-n_1^{^{\prime
}}-n_2+n_2^{^{\prime }}) \\  
&  & -\frac gL\delta _{n_2^{^{\prime }}n_2}\delta (n_1-n_1^{^{\prime
}}+p-p^{^{\prime }})f_1(n_1-n_1^{^{\prime }}-p+p^{^{\prime }}) \\  
&  & -\frac gL\delta _{n_1^{^{\prime }}n_1}\delta (p-p^{^{\prime
}}+n_2-n_2^{^{\prime }})f_1(p-p^{^{\prime }}-n_2+n_2^{^{\prime }})
\end{array}
\end{equation}
where $f_1$ is the function 
\begin{equation}
\label{2.4}f_1(\alpha )=\frac 14\left( 2\delta _{\alpha ,0}+\delta _{\alpha
,2}+\delta _{\alpha ,-2}\right) 
\end{equation}
The center of mass motion is separated by changing coordinates to
\begin{equation}
\label{2.5}
\begin{array}{ccl}
R & = & \frac 1{2+\gamma }\left( x_1+x_2+\gamma y\right)  \\ 
r & = & x_1-x_2 \\ 
\eta  & = & y-\frac{x_1+x_2}2
\end{array}
\end{equation}
Classically the dynamics in the center of mass of the three particles is
ruled by the Hamilton equations
\begin{equation}
\label{2.6}
\begin{array}{ccl}
\stackrel{\bullet }{r} & = & 4p_r \\ 
\stackrel{\bullet }{p_r} & = & -gV^{^{\prime }}(r)+\frac 12gV^{^{\prime
}}(\frac r2-\eta )-\frac 12gV^{^{\prime }}(-\frac r2-\eta ) \\ 
\stackrel{\bullet }{\eta } & = & \frac{2+\gamma }\gamma p_\eta  \\ \stackrel{%
\bullet }{p_\eta } & = & -g\left\{ V^{^{\prime }}(\frac r2-\eta
)+V^{^{\prime }}(-\frac r2-\eta )\right\} 
\end{array}
\end{equation}
That is, the center of mass dynamics is equivalent to the motion of a two
dimensional particle in the potential 
\begin{equation}
\label{2.7}U(r,\eta )=g\left\{ V(r)-V(\frac r2-\eta )-V(-\frac r2-\eta
)\right\} 
\end{equation}
with an effective Hamiltonian 
\begin{equation}
\label{2.8}H_{CM}=2p_r^2+\frac{2+\gamma }{2\gamma }p_\eta ^2+U(r,\eta )
\end{equation}
The potential $U(r,\eta )$ is displayed in Fig.1 with $r$ and $\eta $ in
units of $L$. It has two stable minima at $(r=\frac L3,\eta =0)$ $(r=-\frac
L3,\eta =0)$ and a saddle point at $(r=\eta =0)$. The minima correspond to
configurations with the two positive particles well separated and the
negative one midway between the other two. The saddle point is a collision
state of the two heavy positive particles which however has relatively low
energy because the repulsive energy of the positive particles is compensated
by the attractive interaction with the negative one. In addition to the
unstable fixed point at $r=\eta =0$ there is a whole collection of unstable
periodic orbits along the $r=0$ axis. According to (\ref{1.5}), the energy
of the corresponding saddle point scar is estimated to be 
\begin{equation}
\label{2.9}E_s\simeq -\frac gL+E_{loc}+\frac 12\sqrt{\frac{4g\pi ^2(2+\gamma
)}{\gamma L^3}}
\end{equation}
In case one wants to use scar effects to induce collisions the important
quantity to know is the difference between $E_s$ and the ground state
energy. The structure of the ground state will depend on the order of
magnitude of the physical parameters. For definiteness I assume the ratio of
negative to the positive particles mass to be very small ($O(10^{-4})$) and
also of order $L^{-1}$ in the natural units used to write Eq.(\ref{2.1}).
For the numerical results shown below the values used are $\gamma =2.7\times
10^{-4}$ , $L=13039$ and $g=6$.

From (\ref{2.3}) one sees that to excite the first kinetic mode of the light
particle one needs an energy $\frac{(2\pi )^2}{L^2\gamma }$ as compared to
kinetic energies of order $\frac{(2\pi )^2}{L^2}$ needed to localize the
heavy particles. Therefore one expects the ground state to have a completely
delocalized light particle and a heavy particle relative coordinate
localized around $\frac L2$ . For this configuration both the transversal
localization energy and the potential energy are similar to those for the
scar in (\ref{2.9}). Therefore 
\begin{equation}
\label{2.10}\Delta _s=E_s-E_0\simeq \frac 12\sqrt{\frac{4g\pi ^2(2+\gamma )}{%
\gamma L^3}} 
\end{equation}

A numerical diagonalization of the scaled $LH_1$ Hamiltonian of Eq.(\ref{2.1}%
) was performed on a basis of 729 states restricted to zero total momentum
(center of mass configurations). The structure of the energy spectrum is
shown in Fig.2. The band structure corresponds to the energies needed to
excite the successively higher kinetic modes of the light particle. The
first state in each band lies approximately at the energy%
$$
E_0+n^2\frac{(2\pi )^2}{\gamma L} 
$$
$n=0,1,2,\cdots $ . The ground state energy is $E_0=-5.91$. The amplitude of
the ground state wave function is shown in Fig.3. The amplitude of the state
at the top of the negative energy band ($E_s=-0.905$) is shown in Figs.4a,b.
This is the state that corresponds to the saddle point scar described above.
The difference $E_s-E_0$ is $5.01$ to be compared with the value $5.8$
obtained from the analytical estimate (\ref{2.10}). The state, at the top of
the first band, is, in this band, the one with the largest {\it heavy
particle overlap}, defined as%
$$
\int \left| \Psi (0,\eta )\right| ^2d\eta  
$$
For other bands it also happens that the state with the largest overlap is
at the top of the band. I do not know whether there is a general mechanism
leading to this fact or whether it is a particular feature of this model.
There are in each band other states with large overlaps which however, as a
rule, are not so concentrated along the $r=0$ line (see for example in Fig.5
the contour plot for the state at $E=-1.298$ ).

In this system, the harmonic approximation used in (\ref{1.5}) to predict
the first scar state energy, cannot be reliably carried over to higher
harmonic excitations, because higher energy states are not so localized
around ($r=\eta =0$) and the quadratic approximation for the potential is no
longer valid. For example for the state at the top of the second band
(Fig.6) we have $E-E_0=18.67$ whereas (\ref{1.5a}) with $n=1$ would predict $%
17.4$. For the state at the top of the third band (Fig.7), $E-E_0=51.22$
whereas (\ref{1.5a}) with $n=2$ yields $29.0$.

\section{A three-dimensional periodic Coulomb problem}

The three-dimensional problem considered here is, as before, a periodic
boundary condition many-body problem with two heavy and one light particle
in each lattice cell of volume $L^3$ or, alternatively, a three-particle
system living on the 3-torus. The interaction is the Coulomb potential 
\begin{equation}
\label{3.1}V(|x_1-x_2|)=\frac 1{|x_1-x_2|}
\end{equation}
In problems with periodic boundary conditions and long-range interactions it
would make sense to consider that, acting on the particles in one cell, are
the forces of all the particles in the other cells or at least of those on
the neighboring cells. Alternatively one might change the potential to make
it $L-$periodic. 

Here periodicity is explicitly introduced by the choice of a periodic basis
of momentum eigenstates as in the one-dimensional problem. However for the
computation of the matrix elements of the Hamiltonian only the forces
between the three particles will be considered. This slightly underestimates
the interaction part of the Hamiltonian but it does not change the
qualitative nature of the problem especially in what concerns the small
distance collision effects.

For this problem I will first find the low-lying quantum spectrum and then
see whether the strong overlap states may or may not be interpreted as scar
states. In a basis of momentum eigenstates $\left| \overrightarrow{n_1}%
\overrightarrow{n_2}\overrightarrow{p}\right\rangle $ with $\left|
n\right\rangle =$ $(L)^{-\frac 32}\exp (i\frac{2\pi }L\overrightarrow{n}.%
\overrightarrow{x})$, the matrix elements of the Hamiltonian 
\begin{equation}
\label{3.2}H_2=-\Delta _1-\Delta _2-\frac 1\gamma \Delta
_3+|x_1-x_2|^{-1}-|x_1-y|^{-1}-|x_2-y|^{-1}
\end{equation}
are
\begin{equation}
\label{3.3}
\begin{array}{ccl}
\left\langle \overrightarrow{n_1^{^{\prime }}}\overrightarrow{n_2^{^{\prime
}}}\overrightarrow{p^{^{\prime }}}\left| H_2\right| \overrightarrow{n_1}%
\overrightarrow{n_2}\overrightarrow{p}\right\rangle  & = & \delta _{
\overrightarrow{n_1^{^{\prime }}}\overrightarrow{n_1}}\delta _{%
\overrightarrow{n_2^{^{\prime }}}\overrightarrow{n_2}}\delta _{%
\overrightarrow{p^{^{\prime }}}\overrightarrow{p}}(2\pi )^2\frac
1{L^2}\left( \left| n_1\right| ^2+\left| n_2\right| ^2+\frac{\left| p\right|
^2}\gamma \right)  \\  &  & +\frac 1L\delta _{
\overrightarrow{p^{^{\prime }}}\overrightarrow{p}}^3\delta ^3(%
\overrightarrow{n_1}-\overrightarrow{n_1^{^{\prime }}}+\overrightarrow{n_2}-%
\overrightarrow{n_2^{^{\prime }}})f_2(|\overrightarrow{n_1}-\overrightarrow{%
n_1^{^{\prime }}}-\overrightarrow{n_2}+\overrightarrow{n_2^{^{\prime }}}|)
\\  &  & -\frac 1L\delta _{
\overrightarrow{n_2^{^{\prime }}}\overrightarrow{n_2}}^3\delta ^3(%
\overrightarrow{n_1}-\overrightarrow{n_1^{^{\prime }}}+\overrightarrow{p}-%
\overrightarrow{p^{^{\prime }}})f_2(|\overrightarrow{n_1}-\overrightarrow{%
n_1^{^{\prime }}}-\overrightarrow{p}+\overrightarrow{p^{^{\prime }}}|) \\  &
& -\frac 1L\delta _{\overrightarrow{n_1^{^{\prime }}}\overrightarrow{n_1}%
}^3\delta ^3(\overrightarrow{p}-\overrightarrow{p^{^{\prime }}}+%
\overrightarrow{n_2}-\overrightarrow{n_2^{^{\prime }}})f_2(|\overrightarrow{p%
}-\overrightarrow{p^{^{\prime }}}-\overrightarrow{n_2}+\overrightarrow{%
n_2^{^{\prime }}}|)
\end{array}
\end{equation}
with 
\begin{equation}
\label{3.4}f_2\left( \left| \alpha \right| \right) =\frac 1{2\pi \left|
\alpha \right| }\left( 1-\cos \left( \pi \rho \left| \alpha \right| \right)
\right) 
\end{equation}
and $\rho =2\left( \frac 3{4\pi }\right) ^{\frac 13}$. For the computation
of the spectrum of $H_2$ one considers also states symmetric and
antisymmetric under interchange of the two heavy particles. 
\begin{equation}
\label{3.5}\left| \overrightarrow{n_1}\overrightarrow{n_2}\overrightarrow{p}%
\right\rangle _{\pm }=\frac 1c\left( \left| \overrightarrow{n_1}%
\overrightarrow{n_2}\overrightarrow{p}\right\rangle \pm \left| 
\overrightarrow{n_1}\overrightarrow{n_2}\overrightarrow{p}\right\rangle
\right) 
\end{equation}
with $c=\sqrt{2}$ for $\overrightarrow{n_1}\neq \overrightarrow{n_2}$ and $%
c=2$ for $\overrightarrow{n_1}=\overrightarrow{n_2}$. For the numerical
diagonalization of the Hamiltonian $H_2$ a basis of 3176523 states was
considered. Using momentum conservation and conservation of permutation
symmetry type the matrix is however reduced into invariant blocks to
simplify the computation. Fig.8 shows the lower part of the spectrum for
center of mass (zero total momentum) states with symmetric states denoted by
crosses and antisymmetric ones by dots. In Figs.9a and 9b one compares the
lowest lying antisymmetric $(\Psi _{0a})$ and symmetric $(\Psi _{0s})$
states. The quantity that is plotted is the integrated probability 
\begin{equation}
\label{3.6}\left| \Psi \left( \left| r\right| ,\left| \eta \right| \right)
\right| ^2=\int d\Omega _rd\Omega _\eta \left| \Psi \left( r,\eta \right)
\right| ^2
\end{equation}
The lowest lying symmetric eigenstate displays a strong overlap of the heavy
particles. The nature of this state is better understood from the
two-coordinate projections 
\begin{equation}
\label{3.7}\left| \Psi \left( r_i,\eta _j\right) \right| ^2=\int d^2rd^2\eta
\left| \Psi \left( r,\eta \right) \right| ^2
\end{equation}
the integration being carried over all coordinates other than $r_i$ and $%
\eta _j$. Figs.10a and 10b show $\left| \Psi \left( r_1,\eta _1\right)
\right| ^2$ and $\left| \Psi \left( r_1,\eta _2\right) \right| ^2$. All
other like-coordinate and unlike-coordinate projections are identical to
those shown in the figures.

As in the one-dimensional case studied before, there is a simple relation
between this state and classical orbits. The classical equations for center
of mass motion are
\begin{equation}
\label{3.8}
\begin{array}{ccl}
\stackrel{\bullet }{\overrightarrow{r}} & = & 4
\overrightarrow{p_r} \\ \stackrel{\bullet }{\overrightarrow{p_r}} & = & 
\overrightarrow{r}\left| r\right| ^{-3}-\frac 12\left( \frac{\overrightarrow{%
r}}2-\overrightarrow{\eta }\right) \left| \frac{\overrightarrow{r}}2-%
\overrightarrow{\eta }\right| ^{-3}-\frac 12\left( \frac{\overrightarrow{r}}%
2+\overrightarrow{\eta }\right) \left| \frac{\overrightarrow{r}}2+%
\overrightarrow{\eta }\right| ^{-3} \\ \stackrel{\bullet }{\overrightarrow{%
\eta }} & = & \frac{2+\gamma }\gamma \overrightarrow{p_\eta } \\ \stackrel{%
\bullet }{\overrightarrow{p_\eta }} & = & \left( \frac{\overrightarrow{r}}2-%
\overrightarrow{\eta }\right) \left| \frac{\overrightarrow{r}}2-%
\overrightarrow{\eta }\right| ^{-3}-\left( \frac{\overrightarrow{r}}2+%
\overrightarrow{\eta }\right) \left| \frac{\overrightarrow{r}}2+%
\overrightarrow{\eta }\right| ^{-3}
\end{array}
\end{equation}
For collisions or near-collisions to take place at low energies the light
particle must be near the two heavy particles part of the time in order for
the attractive energy to compensate the strong repulsion. This implies some
localization of the light particle in one dimension at least. Localization
in more dimensions however is very costly in kinetic energy. Therefore it is
likely to find, associated to the low lying symmetric state, orbits
corresponding to planar motion but not to collinear motion. From (\ref{3.8})
it follows that there are orbits in the plane ($r_1,\eta _2$), that is, if
at $t=0$ one has $r_2=p_{r2}=r_3=p_{r3}=\eta _1=p_{\eta 1}=\eta _3=p_{\eta
3}=0$ then the same holds true for all $t$. The phase-portrait in the plane (%
$r_1,\eta _2$) is symmetric about the $r_1=0$ line. In Fig.11a and 11b I
have plotted two typical orbits. Periodic boundary conditions are imposed at 
$\frac 12$ and $-\frac 12$. In-between the orbits that move to the right and
to the left there is the separatrix at $r_1=0$ that passes through the
singular point of the potential. It is precisely along this separatrix that
the state $\Psi _{0s}$ is strongly scarred. In this case we have not found a
saddle scar in the sense defined in the introduction because the separatrix
goes through a singular point of the potential. It is possible that by a
change of time, and using the well-known regularization\cite{Kustaa} of the
Kepler problem one may still be able to formally interpret the state as a
saddle scar. This is not however very important. What is important to notice
is that once again we have found that the low-energy collisional state is
associated to an unstable feature of the classical phase-portrait.

\section{Conclusions}

Through the scar effect, orbits that correspond to zero measure classical
configurations may be reached and stabilized in quantum mechanics by
resonant excitation with the appropriate energy. Their location in the
energy spectrum may, in favorable cases, be found either from theoretical
considerations (from symmetry, from being at the top of bands, etc.) or from
absorption experiments.

This characteristically quantum phenomenon, may be practically used to
induce reactions which are favored by unstable or difficult to reach
configurations. An obvious potential application is to fusion reactions.
Most practical nuclear fusion mechanisms proposed so far have a
two-mechanism nature. For example muon catalyzed fusion relies on the fact
that the muon is 200 times heavier than the electron to bring the bound
nuclei together, but then it is the tunneling effect than will eventually
allow the nuclei to fuse. In toroidal plasma confinement it is the magnetic
field configuration that keeps the ions together, but then it is radio
frequency or ion injection heating that supplies them with enough kinetic
energy to overcome the Coulomb barrier. What I am proposing here is that if
enough deuterons, for example, are confined in a periodic medium (a crystal
lattice, for example) then, resonant excitation and the scar effect may be
used to excite collisional states and induce fusion reactions. I strongly
emphasize that like in the known fusion methods we should separate the two
problems of confinement and collision. The crystal lattice only serves as a
confinement device, an additional collision mechanism being needed, which
the scar effect, discussed in this paper, may provide. This is contrary to
the hopes of the cold fusion saga where spontaneous fusion was expected just
from confining deuterons in a lattice. Actually a simple calculation shows
that in the problem discussed in Section 3 the separation between the ground
state and the first scar state is such that thermal excitation is highly
improbable. However resonant excitation by electromagnetic radiation seems
possible.

For other applications of the scar effect and in particular of its saddle
point enhancement one might think of catalyzing chemical reactions on
lattice substrates.

In conclusion: unstable configurations corresponding to unstable periodic
orbits are in classical mechanics of little use unless very sophisticated
control methods are used\cite{Ott}. This is because, for each energy level,
the stable invariant measures are smoothly distributed all over the energy
surface. In this sense quantum mechanics is more parsimonious because for
each energy level it displays just a fraction of the blurred picture of
smooth classical dynamics. In particular, by isolating through the scar
effect improbable classical configurations, quantum mechanics is, for
practical applications, a cure to the classical mechanics curse of
ergodicity on the energy surface.

\section{Figure captions}

Fig.1 - Effective center of mass potential for the one-dimensional problem

Fig.2 - Energy spectrum in the one-dimensional problem

Fig.3 - Ground state wave function

Figs.4a,b - Wave function of the state at the top of the negative energy band

Fig.5 - Contour plot for the state at energy $E=-1.298$

Fig.6 - Wave function of the state at the top of the second energy band

Fig.7 - Wave function of the state at the top of the third energy band

Fig.8 - Energy spectrum in the three-dimensional problem

Figs.9a,b - $|\Psi (|r|,|\eta |)|^2$ for the lowest lying antisymmetric and
symmetric states

Fig.10a - $|\Psi (r_1,\eta _1)|^2$ for the lowest lying symmetric state

Fig.10b - $|\Psi (r_1,\eta _2)|^2$ for the lowest lying symmetric state

Figs.11a,b - Two classical center of mass orbits in the plane $(r_1,\eta _2)$

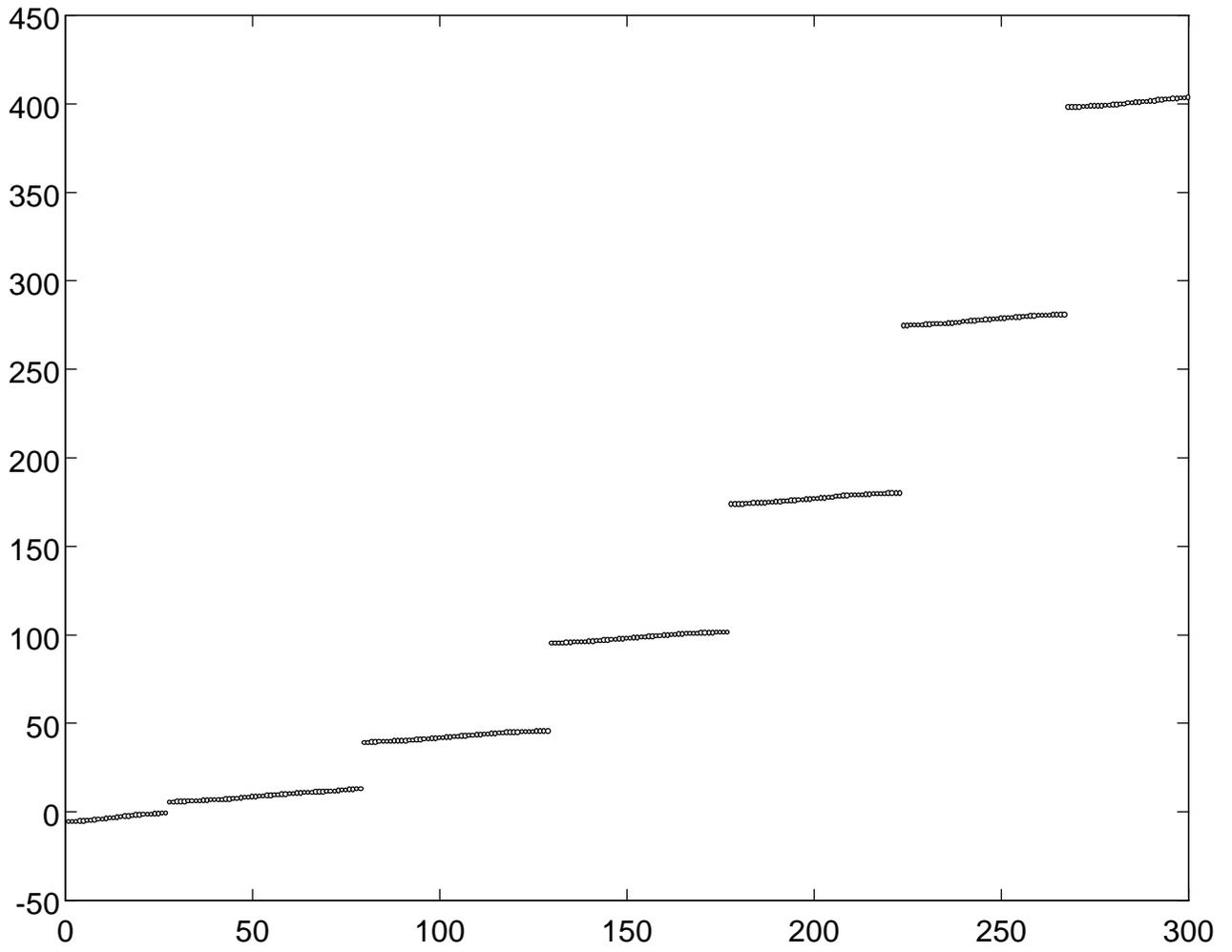
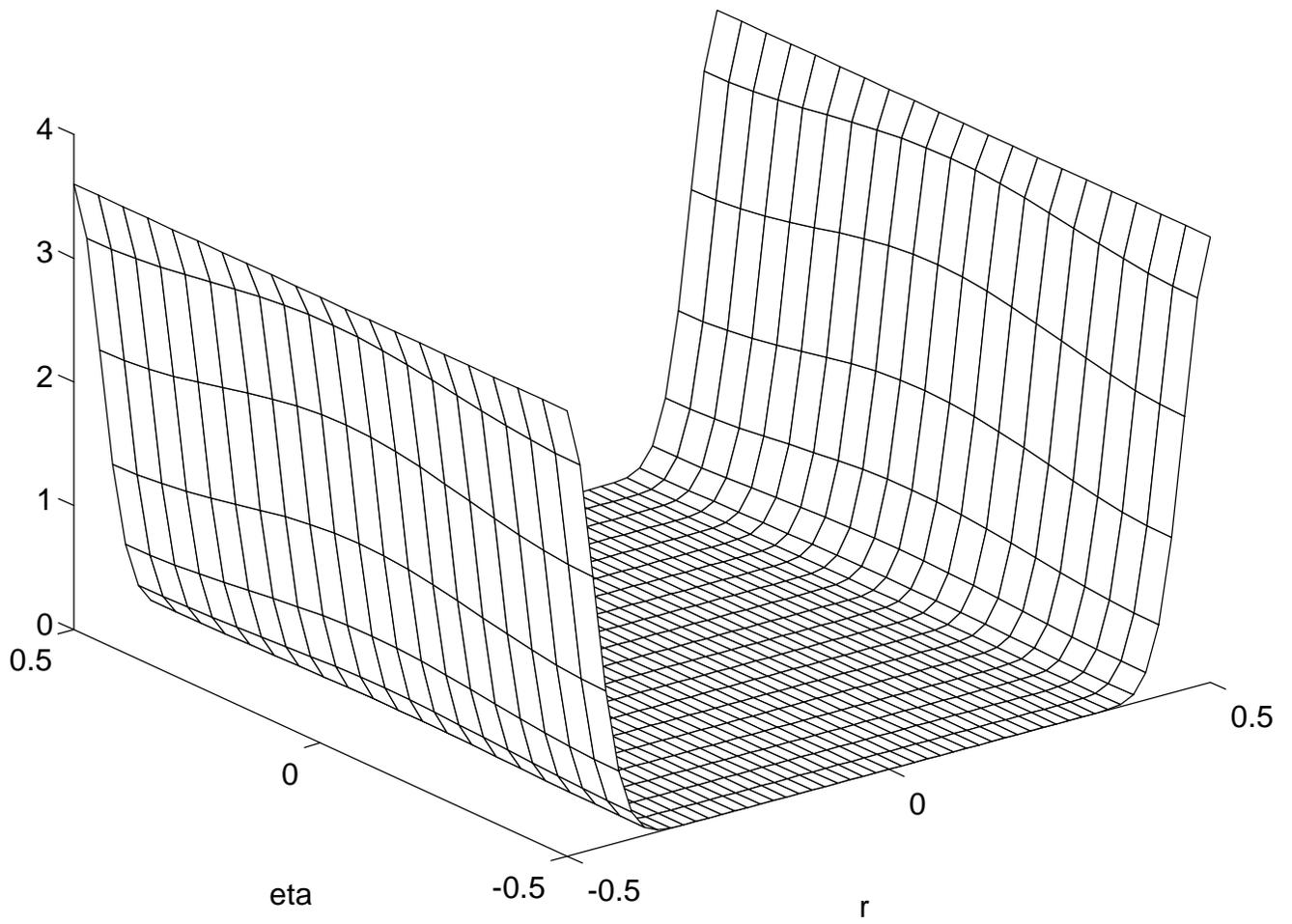
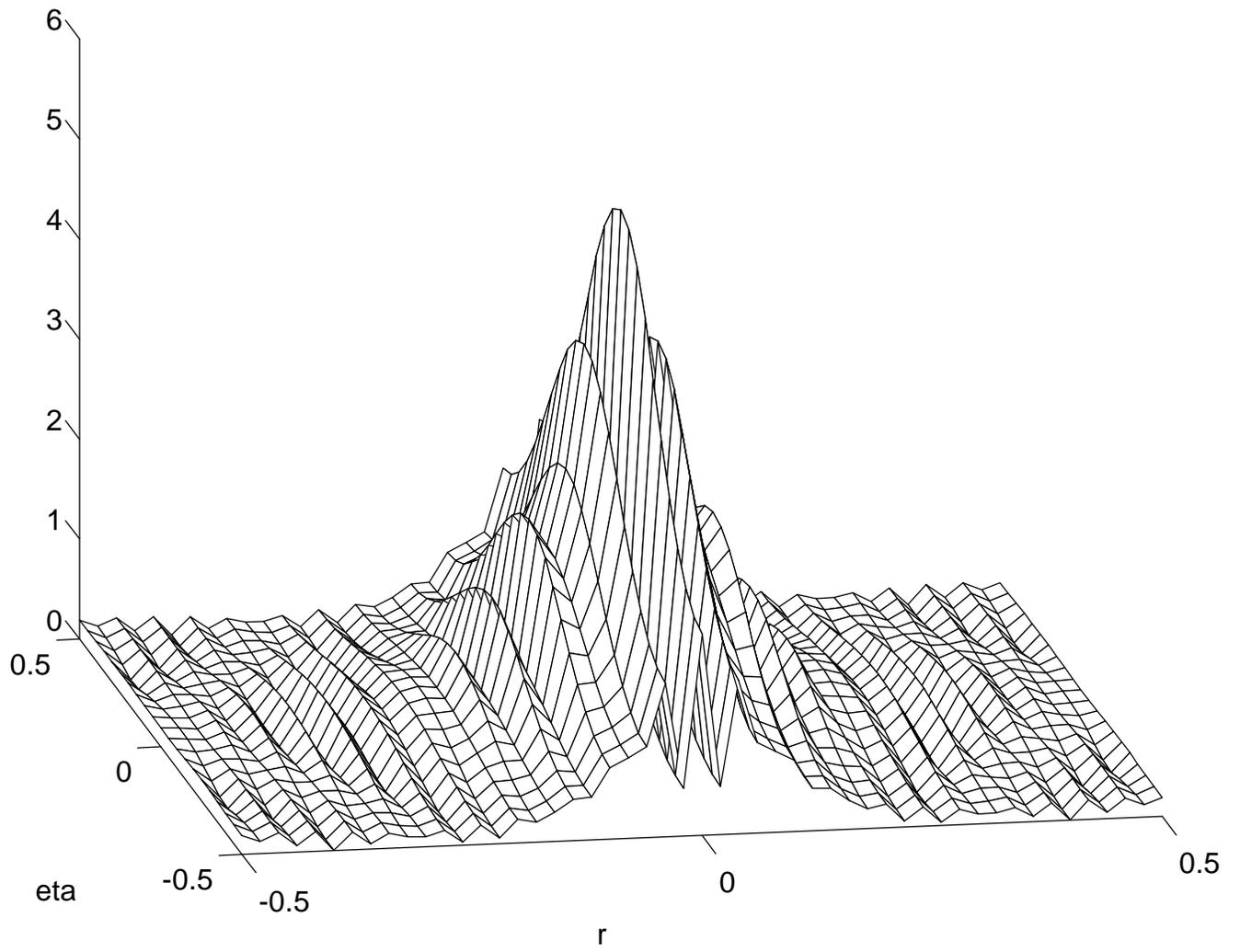
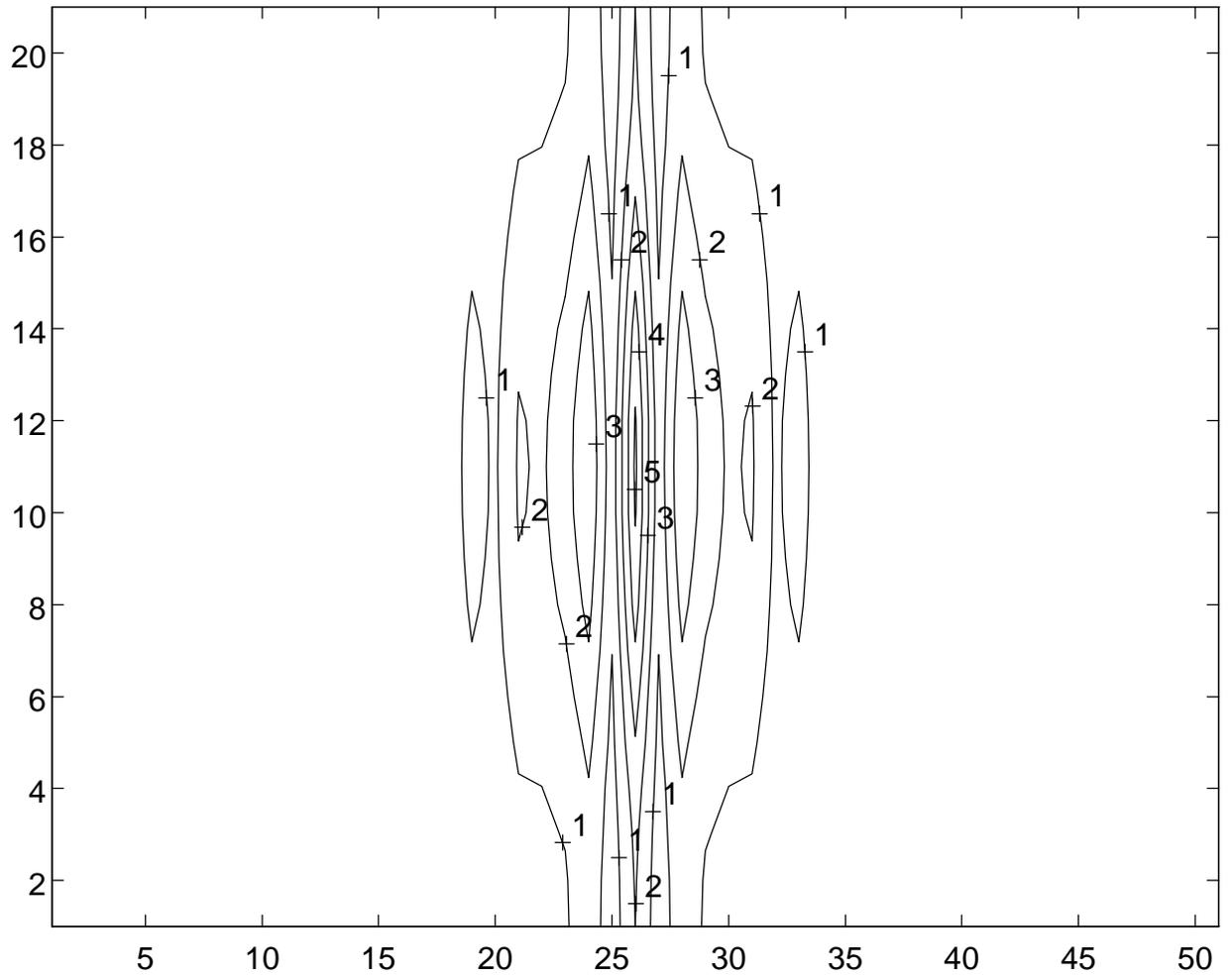
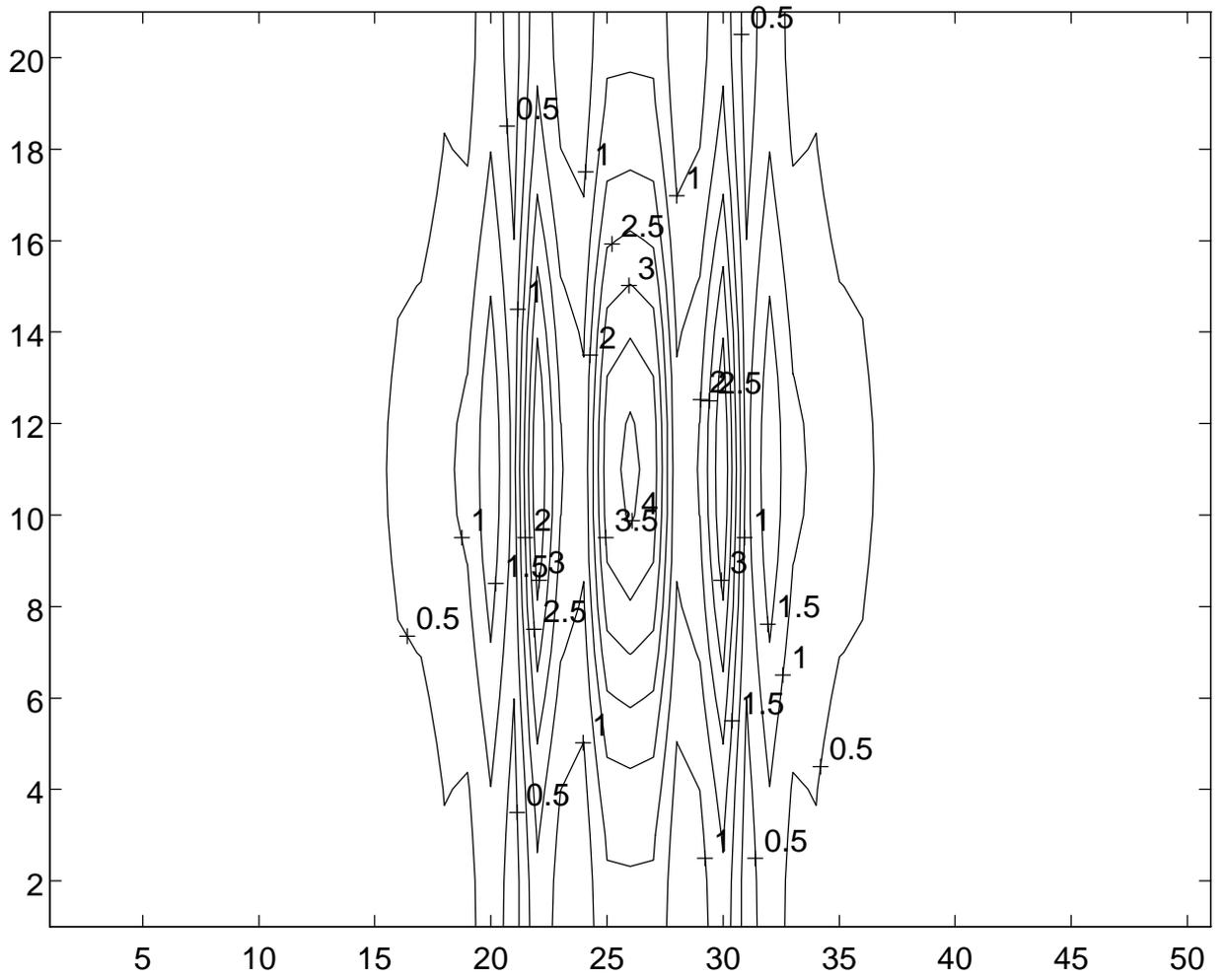
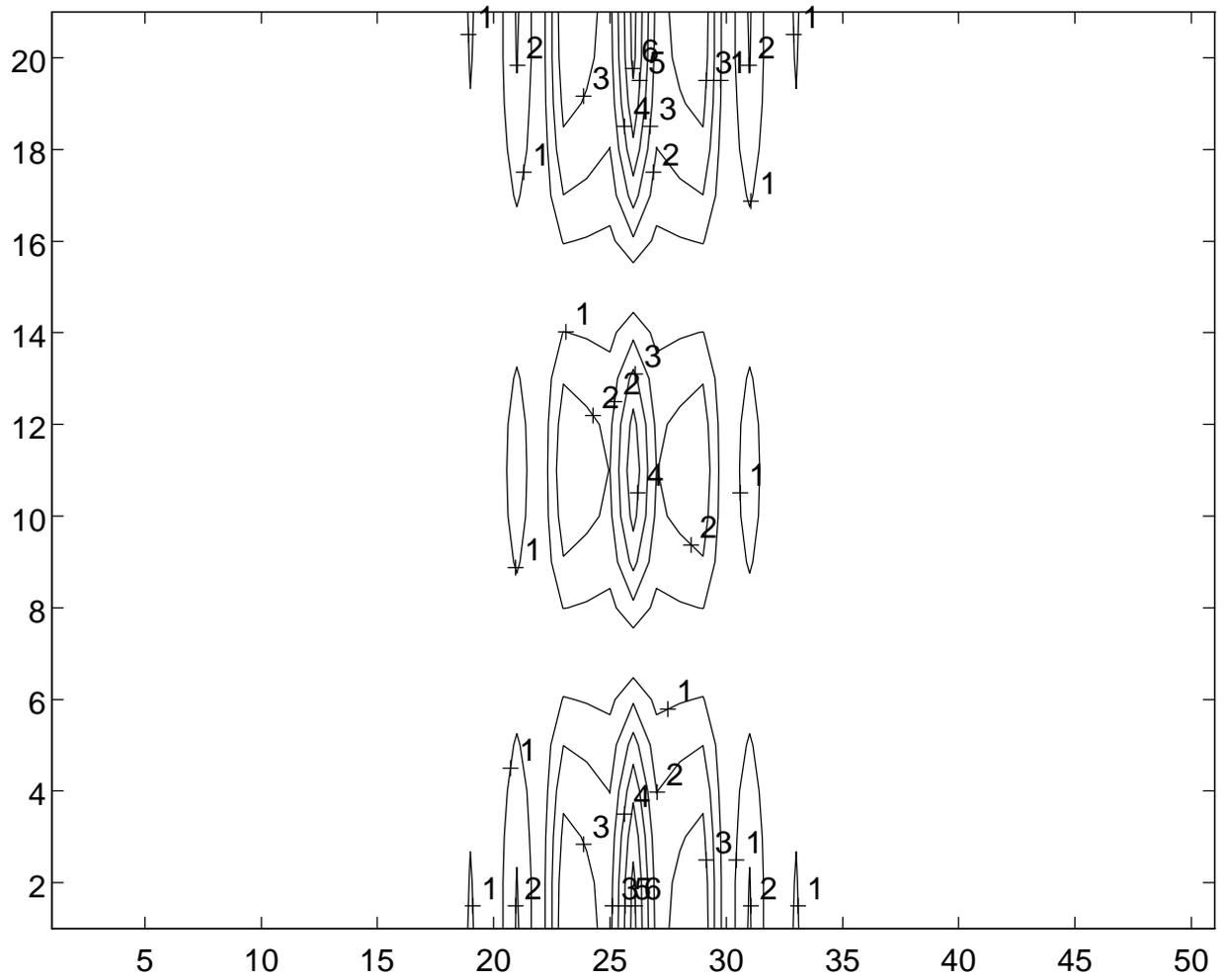
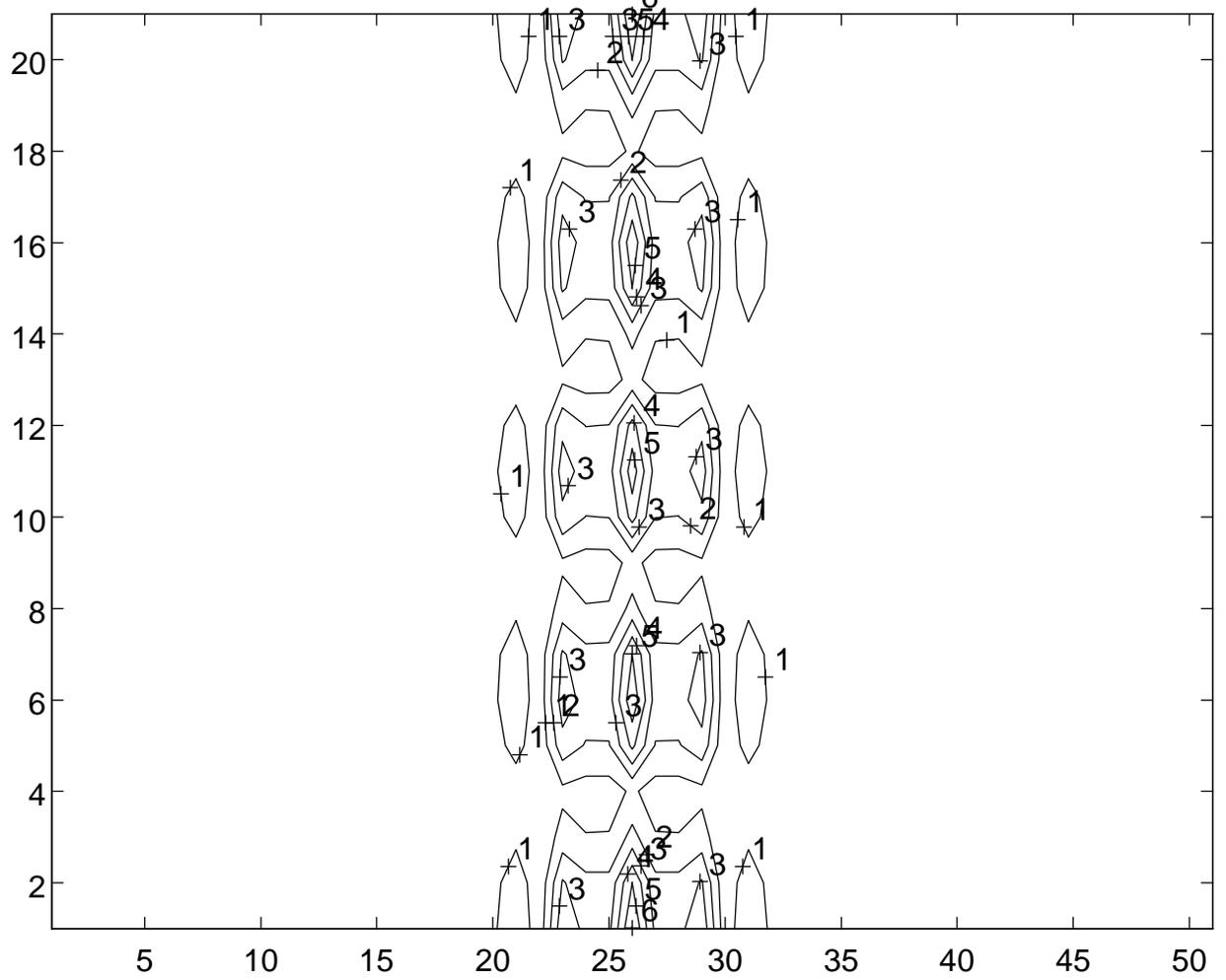
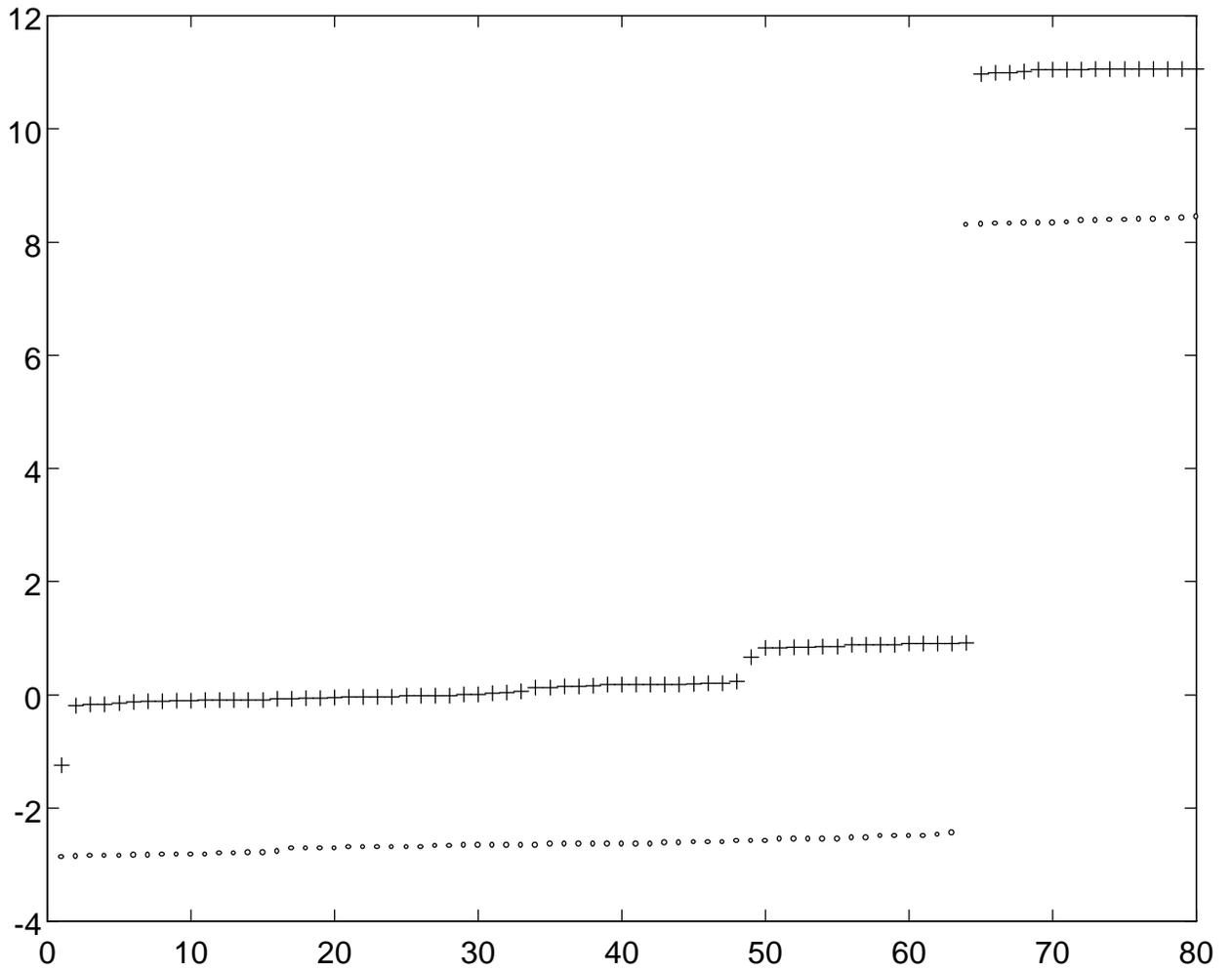
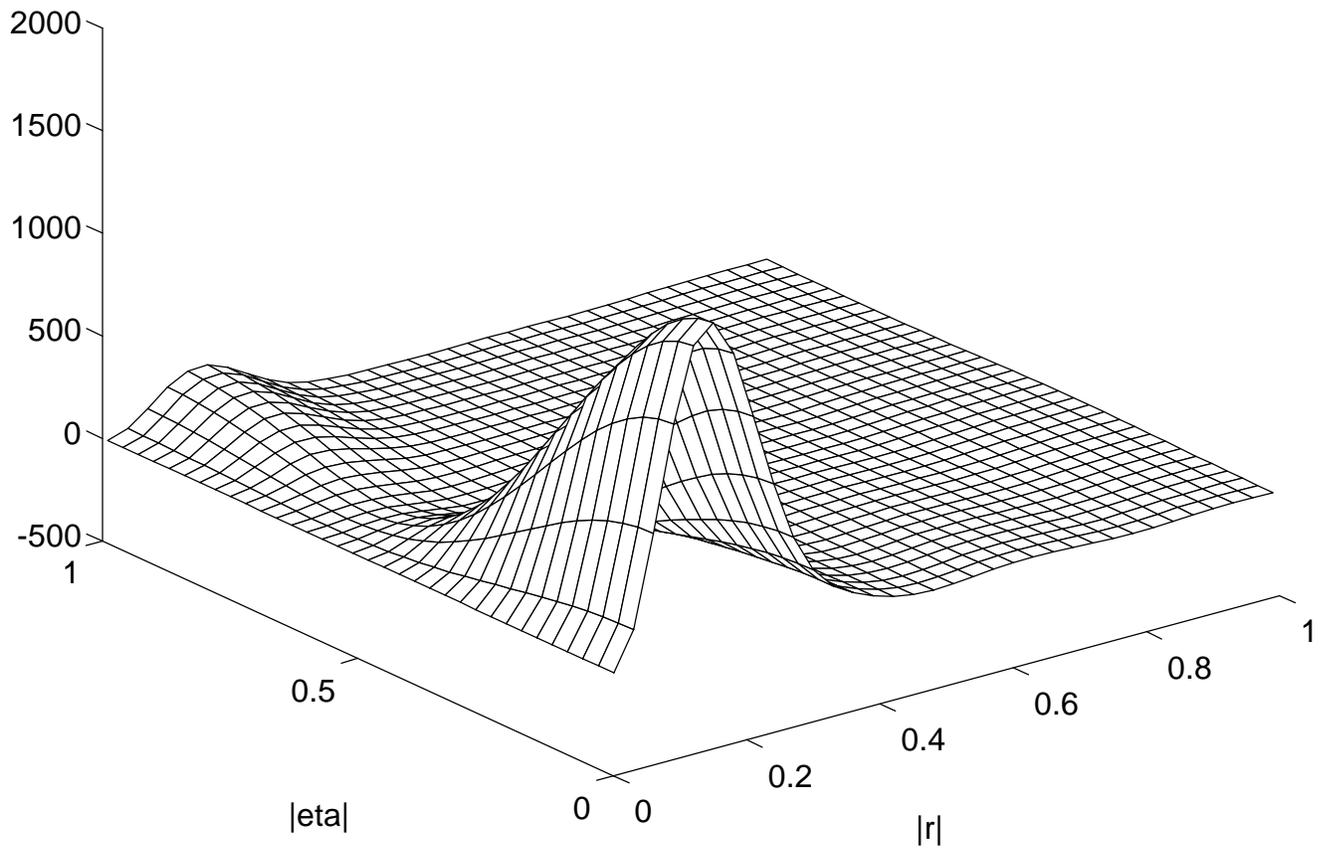
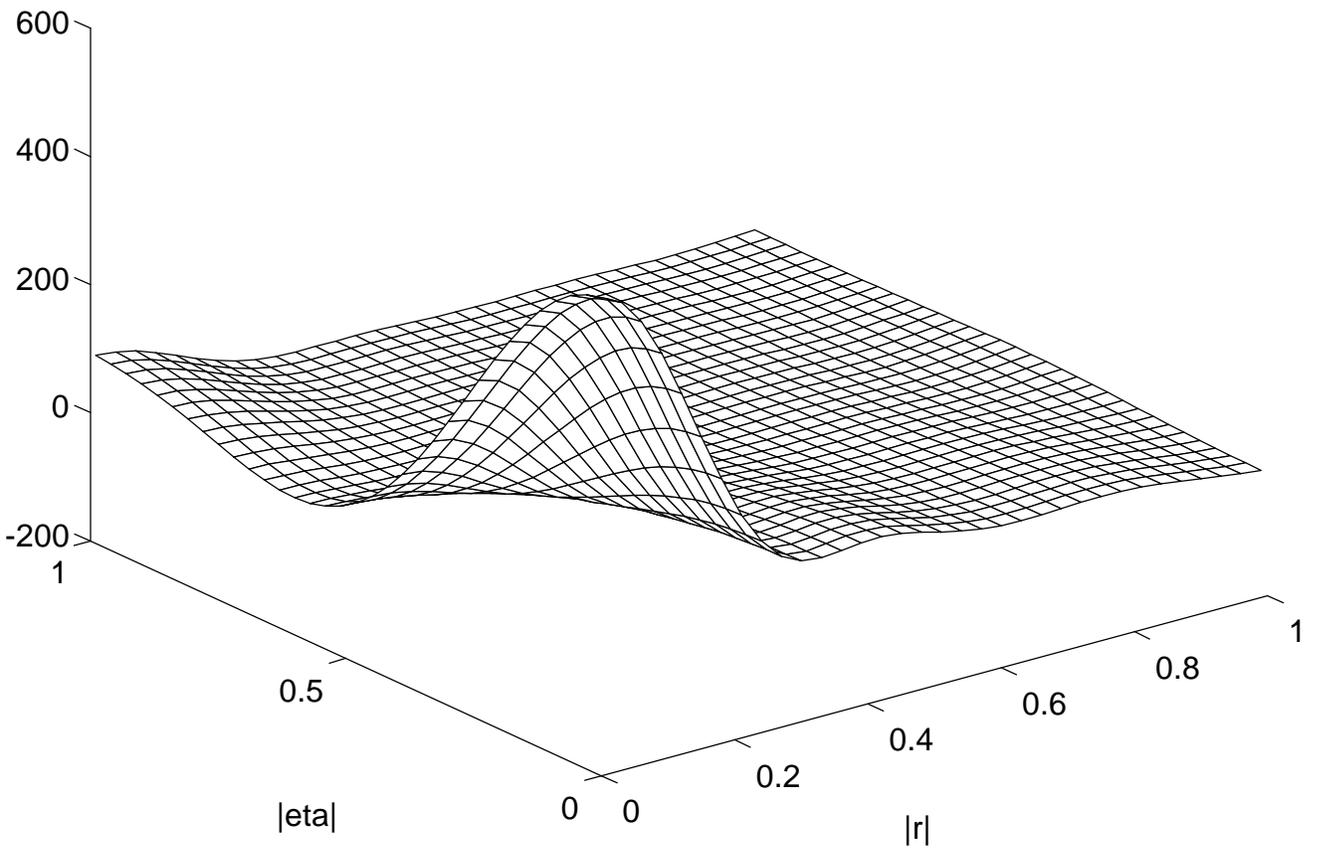
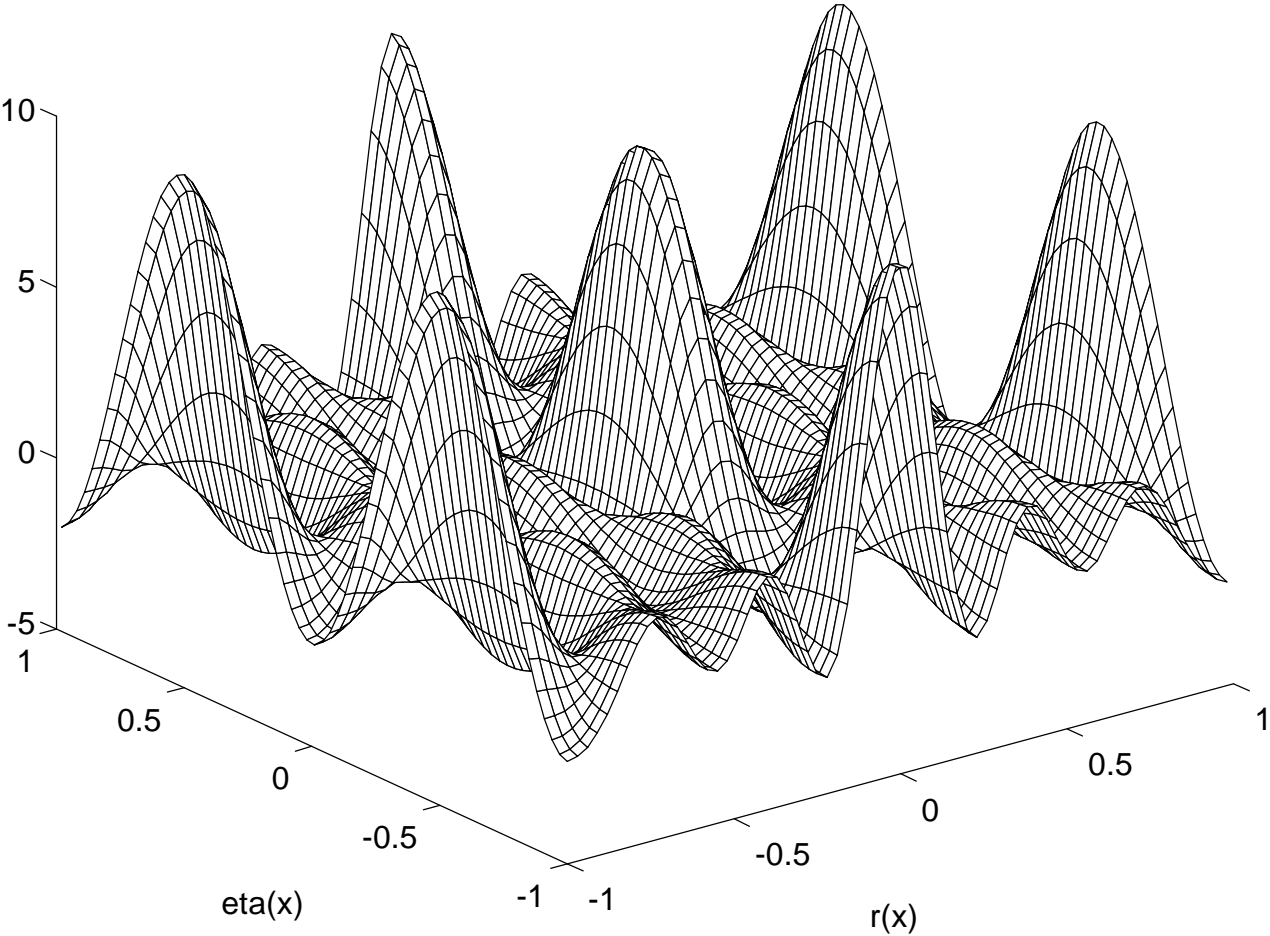
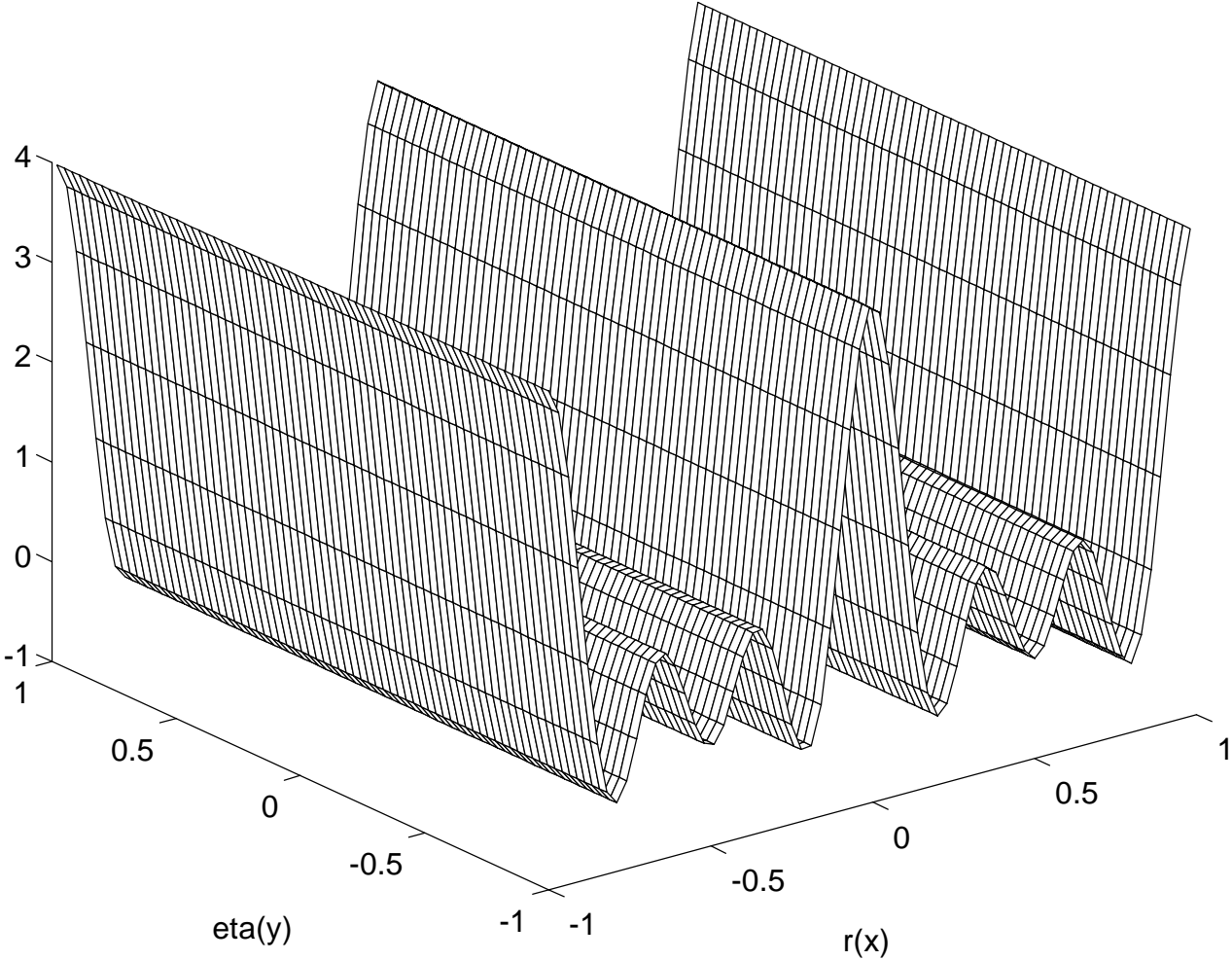
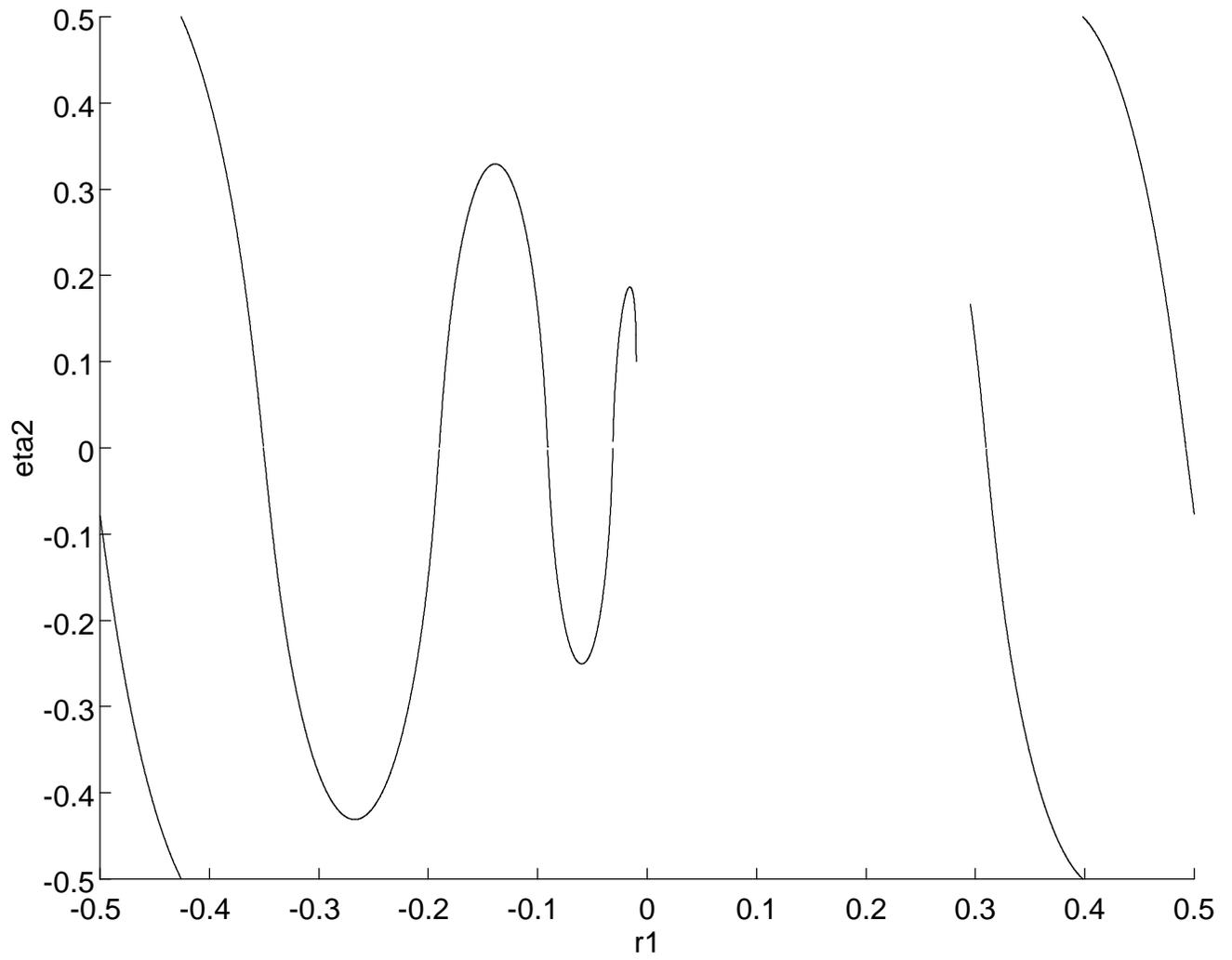
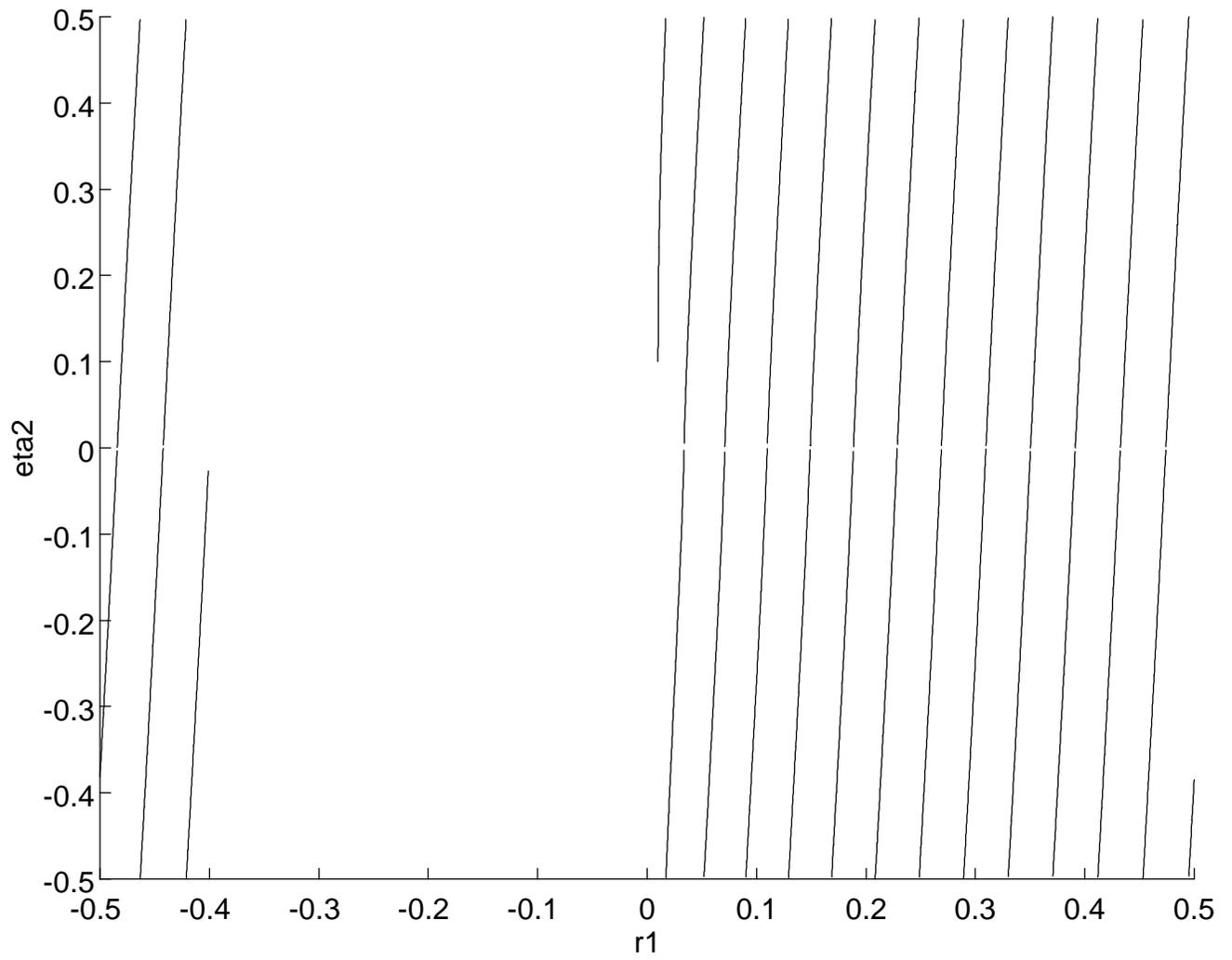
\end{document}